# High-Performance Logic and Memory Devices Based on a Dual-Gated MoS$_2$ Architecture

Fuyou Liao,[†, ‡] Zhongxun Guo,[†, ‡] Yin Wang,[†, ‡] Yufeng Xie,[†, ‡] Simeng Zhang,[†] Yaochen Sheng,[†] Hongwei Tang,[†] Zihan Xu,[&] Antoine Riaud,[†] Peng Zhou,[†,*] Jing Wan,[§] Michael S. Fuhrer,[⊥,‖], Xiangwei Jiang,[#] David Wei Zhang,[†] Yang Chai,[‡,*] Wenzhong Bao[†,*]

[†] State Key Laboratory of ASIC and System, School of Microelectronics, Fudan University, Shanghai 200433, China.

[‡] Department of Applied Physics, The Hong Kong Polytechnic University, Hong Kong, China.

[§] State Key Laboratory of ASIC and System, School of Information Science and Engineering, Fudan University, Shanghai 200433, China.

[⊥] ARC Centre of Excellence in Future Low-Energy Electronics, Monash University, 3800 Victoria, Australia.

[‖] School of Physics and Astronomy, Monash University, 3800 Victoria, Australia.

[#] Institute of Semiconductors, Chinese Academy of Sciences, Beijing 100083, China.

[&] Six carbon Tech. Shenzhen, Shenzhen 518106, China.

**ABSTRACT:** In this work, we demonstrate a dual-gated (DG) MoS$_2$ field effect transistors (FETs) in which the degraded switching performance of multilayer MoS$_2$ can be compensated by the DG structure. It produces large current density (>100 μA/μm for a monolayer), steep subthreshold swing (*SS*) (~100 mV/dec for 5 nm thickness), and high on/off current ratio (greater than 10$^7$ for 10 nm thickness). Such DG structure not only improves electrostatic control but also provides an extra degree of freedom for manipulating the threshold voltage ($V_{TH}$) and *SS* by separately tuning the top and back gate voltages, which are demonstrated in a logic inverter. Dynamic random access memory (DRAM) has a short retention time because of large OFF-state current in the Si MOSFET. Based on our DG MoS$_2$-FETs, and a DRAM unit cell with a long retention time of 1260 ms are realized. A large-scale isolated MoS$_2$ DG-FETs based on CVD-synthesized continuous films is also demonstrated, which shows potential applications for future wafer-scale digital and low-power electronics.

1. INTRODUCTION

Recently, two-dimensional layered transition metal dichalcogenides (TMDs) have attracted significant fundamental research attention[1-2] and MoS$_2$ is one prominent representative.[3] Unlike graphene, which lacks a bandgap, MoS$_2$ has a direct gap of 1.8 eV [3-4] for a monolayer, producing high on/off current ratio ($I_{ON}/I_{OFF}$ >10$^8$) [5-6] and outstanding optoelectronic properties.[7] As a channel material for field-effect transistors (FETs), the ultra-thin body also helps suppress short channel effects (SCE),[8] and the lower dielectric constant of MoS$_2$ ($\varepsilon_{MoS_2} = 6.8 \sim 7.1$) compared to that of Si ($\varepsilon_{Si}$=11.9) can further suppress SCE as the characteristic length $L \propto \varepsilon^{0.5}$.[9] Bilayer (BL) or multilayer (ML) MoS$_2$ are also promising due to their smaller bandgap and higher mobility, yielding larger driving current.[10-12] Therefore, tuning the thickness of MoS$_2$ provides greater flexibility for use in various applications, including high speed transistors, [13] nonvolatile memory,[14-15] ultrasensitive photodetector,[7, 16-17] and integrated circuits.[18-19] However, FETs based on BL or ML-MoS$_2$ suffer from degradation of a larger subthreshold swing (*SS*) and a lower $I_{ON}/I_{OFF}$ value [5] due to increased OFF-state current. It is mainly attributed to the single gate structure and the lack of doping strategies to form p-n junctions at the source and drain regions. The increasing OFF-state current also limits its application in low power devices and dynamic memories. On the other hand, as the device size keeps shrinking, manipulating and separating the threshold voltage ($V_{TH}$) becomes increasingly difficult due to issues surrounding the ever-challenging doping and gate deposition processes. [20]

Using a dual-gated (DG) transistor structure is a natural solution, similar to that in Fin-FET technology,[21-22] as it provides better electrostatic control over the channel region and is easier to fabricate in 2D-TMDs. DG-FETs also allow both gates to be tuned separately, thus providing an additional degree of freedom for separating and fine-tuning $V_{TH}$ in each circuit block. Furthermore, a local DG structure takes this advantage to its extreme and provides control over $V_{TH}$ in every single FET in the circuit. For 2D-TMDs, most previously studied devices include a high-*k* dielectric layer top gate (TG) structure with a global back gate (BG).[23-24] The TG dielectric is typically a thin layer (~30 nm) of high-*k* HfO$_2$,[5, 25-26] while the BG is often a 300 nm thick SiO$_2$ layer.[6, 27] Recently, Liu *et al.*[28] fabricated a small footprint DG MoS$_2$ FET by global BG and local TG with Al$_2$O$_3$/HfO$_2$ as BG dielectric and 10 nm BN as TG dielectric. Such a combination of a global BG and local TG can naturally form a DG structure, and the process for fabricating the device is fairly straightforward. A summary of exfoliated MoS$_2$ DG-FETs in the literature is seen in table S1. However, such an asymmetric device structure does not retain the advantages of a DG-FET; when the BG and TG are both active with similarly applied voltages, the channel is primarily controlled by the TG because the

BG has much lower dielectric capacitance. It is also difficult to integrate it in large-scale circuits as the BG is global and cannot be controlled separately for a single FET.

In this study, a DG architecture with symmetric BG and TG is developed for MoS$_2$ FETs. Compared to typical single gated (SG) MoS$_2$ FETs, such a DG structure provides independent gate control for both local BG and TG, and a high-*k* dielectric is used in both gates. $V_{TH}$ and *SS* can be modulated by tuning the BG and TG separately, providing an excellent channel current ($I_D$) modulation when the BG and TG are electrically connected, similar to the Fin-FET or gate-all-around (GAA) techniques. We then take advantage of the DG structure in a MoS$_2$ FET to operate a logic inverter in different working modes. A one transistor one capacitor (1T1C) dynamic random access memory (DRAM) unit cell with ultra-long retention time is also presented herein. Finally, we illustrate a practical application, where a die with isolated 81 MoS$_2$ DG-FETs based on a CVD-synthesized MoS$_2$ continuous film is fabricated. Such device architecture can be extended to other TMDs and 2D layered materials, thus providing a new paradigm for this class of devices.

## 2. RESULTS AND DISCUSSION

Figure 1a shows a cross-sectional schematic of MoS$_2$ DG-FET, where the source/drain electrodes form top contacts with an exfoliated MoS$_2$ sheet. The channel length *L* is 2~4 μm and a small intentional overlap between the source/drain and two gate electrodes was designed to provide thorough channel control. A detailed device fabrication procedure is described in the Methods section and Supplementary Figure S1 and S2. Figure 1b shows an optical microscope image of the as-fabricated monolayer MoS$_2$ DG-FET. Both BG and TG structures include a high-*k* dielectric layer (15 nm thick HfO$_2$) deposited using ALD at 180°C. Unlike the gate-first technique in the BG configuration,[29] the

TG requires a 2-nm-thick $Y_2O_3$ seeding layer to ensure uniform deposition of $HfO_2$. The cross-sectional transmission electron microscopic (TEM) of a $MoS_2$ DG-FET is presented in Figure S3, which exhibits a uniform and compact interface between $MoS_2$ and $HfO_2$ with symmetric DG structure.

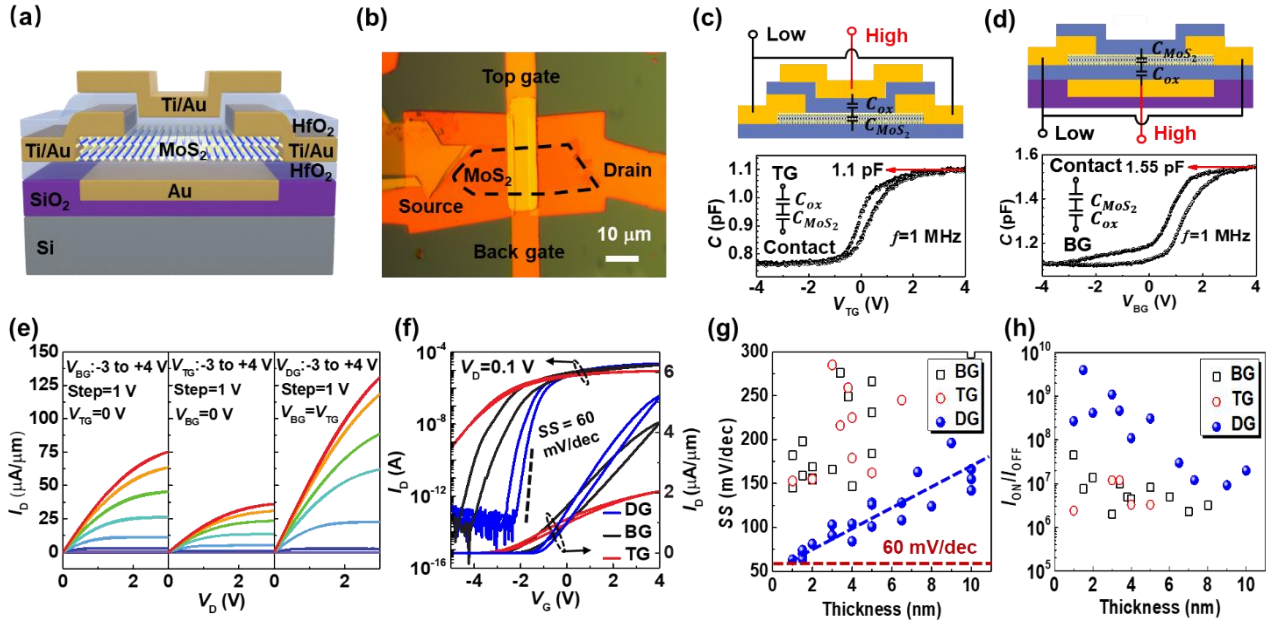

**Figure 1.** (a) Three-dimensional schematic of a $MoS_2$ DG-FET. (b) Optical image of an as-fabricated monolayer $MoS_2$ DG-FET. The scale bar is 10 μm. (c-d) Capacitance-voltage curves (below) and measurement connection (above) in the TG- and BG modes. (e) $I_D$-$V_D$ output curves with forward and backward sweep for various applied $V_{BG}$ (left), $V_{TG}$ (middle), and $V_{DG}$ (right) values (all range from −3 to +4 V with 1 V increments). The thickness of $MoS_2$ is nearly 1 nm. (f) $I_D$-$V_G$ transfer curves by forward and backward sweeping $V_{BG}$, $V_{TG}$, and $V_{DG}$ with $V_D$=0.1 V. (g) $SS$ extracted from the BG (black hollow squares), TG (red hollow circles), and DG (blue solid circles) transfer curves for $MoS_2$ with different thicknesses. The dashed line is 60 mV/dec and the blue line is the fitting result. (h) Dependence of $I_{ON}/I_{OFF}$ in DG-FETs on the thickness of $MoS_2$.

We further perform $C$-$V$ measurements to evaluate the quality of the two resulting interfaces between $HfO_2$ and $MoS_2$ sheet and measure the gate oxide capacitance, as shown in Figure 1c and 1d. During the measurement, the source and drain electrodes are grounded, while a voltage bias is applied to the

TG (BG). High-frequency *C–V* curves at 1 MHz show a clear transition from accumulation to depletion, similar to that observed in a typical n-type MOS capacitor. Moderate hysteresis and low leakage current (below 10 fA) are observed in ambient conditions, illustrating the high quality of the dielectric layer. Moreover, in a typical MoS$_2$ FET structure, two different capacitances are considered in series: the capacitance of the gate oxide HfO$_2$ ($C_{ox}$) and the capacitance of the depleted MoS$_2$ channel ($C_{MoS_2}$). An equivalent circuit is shown in the inset of Figure 1c and d. The effective oxide capacitance for both BG and TG can be extracted from *C-V* curves as follows. The total capacitance is $\frac{1}{C_{total}} \approx \frac{1}{C_{ox}} + \frac{1}{C_{MoS_2}}$, hence the MoS$_2$ channel is under accumulation mode when a positive gate voltage is applied, resulting in $C_{total} \approx C_{ox}$. Thus, $C_{ox}$ of the HfO$_2$ dielectric can be extracted. The gate capacitance of the TG is estimated to be 1.1 pF, which is slightly lower than that of the BG (1.55 pF), partially due to the extra Y$_2$O$_3$ seeding layer. These measured results correspond to the calculated result of TG (1.0 pF) and BG (1.59 pF) capacitance. (see Figure S4 for details)

Our MoS$_2$ DG-FET can operate in three modes, *1*) TG mode which sweeps $V_{TG}$ with fixed $V_{BG}$. *2*) BG mode which sweeps $V_{BG}$ with fixed $V_{TG}$, and *2*) DG mode which sweeps TG and BG voltages simultaneously. Figure 1e shows the drain current $I_D$ as a function of drain voltage $V_D$ for a MoS$_2$ DG-FET biased with various $V_{BG}$, $V_{TG}$, and $V_{DG}$ values. These output characteristics for the BG and TG show typical n-MOSFET behavior, with an ON-state saturation current $I_D$ = 75 µA/µm and 36 µA/µm for the BG and TG modes, respectively. The BG mode exhibits a slightly stronger current modulation capability than that of TG mode, which can be attributed to the extra seeding layer and larger contact resistance ($R_c$) in the TG mode. The DG mode exhibits even stronger electrostatic control than either the BG or TG mode when driven with the same gate voltage, reaching up to 130 µA/µm. Figure 1f represents transfer curves for the same device. Compared to the device with single BG or TG gating, a steeper switching can be observed under the DG mode, with *SS* approaching the thermionic limit of 60 mV/dec. Figure 1g shows *SS* as a function of thickness for a batch of DG devices with different thicknesses. The thickness was measured using atomic force microscopy (AFM) and Raman spectroscopy (see Supplementary Figure S5 and S6). These results illustrate the level of electrostatic control one can exert over the channel in the DG architecture; the *SS* values under the DG mode are all smaller than those measured in the SG (TG or BG) situation. The *SS* is less than 100 mV/dec when the $MoS_2$ thickness is smaller than 5 nm. More surprisingly, *SS* ~ 150 mV/dec is achieved by the DG-mode for a 10 nm thick $MoS_2$ device. Under such thickness, it is normally difficult to gate a $MoS_2$ FET with a single gate (see Supplementary Figure S7), and this is comparable to that achieved in an SG 1L-$MoS_2$ FET. Figure 1h shows that thinner $MoS_2$ FETs have higher $I_{ON}/I_{OFF}$ values, and $I_{ON}/I_{OFF}$ can also be improved under the DG mode for $MoS_2$ films with various thicknesses.

To further investigate the gate modulation mechanism in the DG architecture, $I_D$ as functions of both $V_{TG}$ and $V_{BG}$ is displayed in a 2D diagram shown in Figure 2a. When $V_{TG}$ = −4 to 0 V, it shows parallel diagonal contours with a constant current (tangent slope≈1), indicating that both gates provide similar modulation capability, thanks to the symmetric gate stack ($C_{BG}/C_{TG}$≈1). Meanwhile, when $V_{TG}$ = 0 to 4 V, the contours are nearly vertical, indicating that $I_D$ is now mainly modulated by the BG. This intriguing phenomenon can be attributed to the imperfect symmetric device architecture (e.g., top-contacted electrodes), where $V_{TG}$ only modulates the channel region while $V_{BG}$ acts on the contact and the channel region[26, 30]. This will be explained in further detail later.

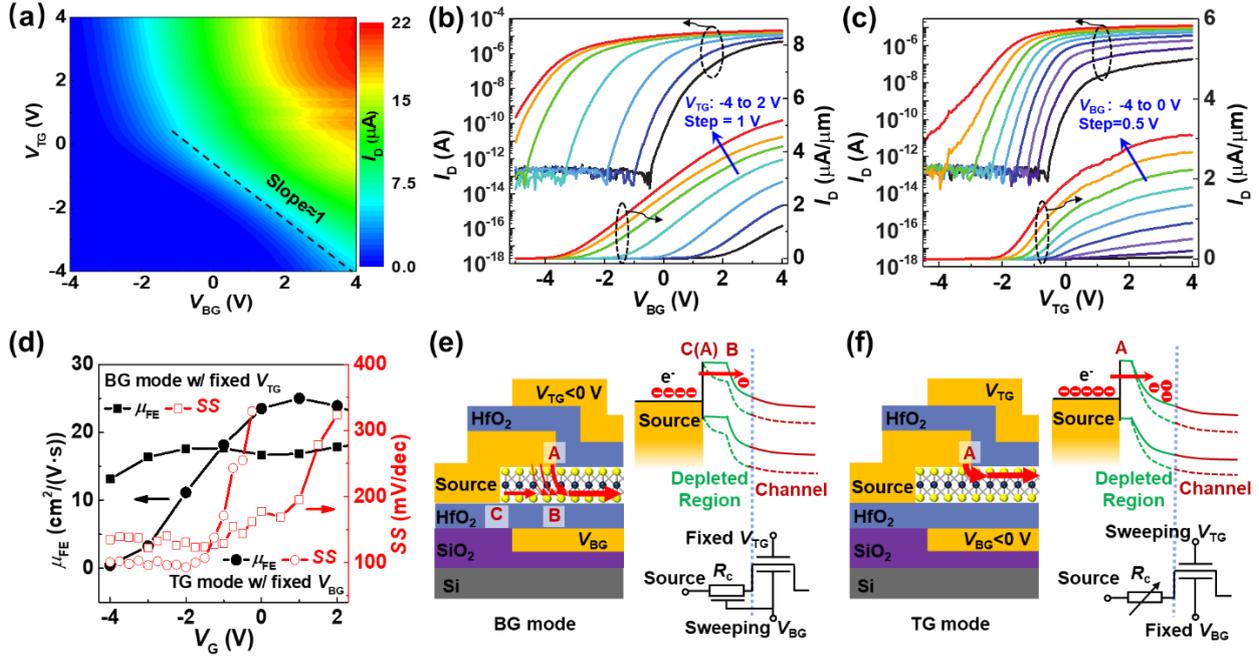

**Figure 2.** Electrical characteristics of a monolayer MoS$_2$ DG-FET. (a) 2D mapping of $I_D$ as functions of $V_{TG}$ and $V_{BG}$ at a constant $V_D = 0.1$ V. (b) BG Transfer characteristics with $V_{TG}$ ranging from −4 to 2 V in 1 V increments. (c) TG transfer characteristics with $V_{BG}$ ranging from −4 to 0 V in 0.5 V increments. (d) The extracted $\mu_{FE}$ (black symbols) and $SS$ (red symbols) operated under the two modes. Square symbols are extracted from curves in (b), corresponding to the BG mode under various $V_{TG}$, and circular symbols are extracted from curves in (c), corresponding to the TG mode under various $V_{BG}$. (e) A cross-sectional view of the contact region when operated under BG mode with $V_{TG}<0$ V (left), and the corresponding energy band diagrams for BG modulation (upper right) and equivalent circuit diagram (lower right). (f) A cross-sectional view of the contact region when operated under TG mode with $V_{BG}<0$ V (left), and the corresponding energy band diagrams for TG modulation (upper right) and equivalent circuit diagram (lower right). The solid and dashed lines in the band diagrams correspond to the negative and positive gate voltages, respectively.

More detailed BG transfer characteristics with varying $V_{TG}$ values were extracted from Figure 2a is shown in Figure 2b. This figure illustrates that the MoS$_2$ FET under BG mode exhibits a typical n-type FET transfer characteristics curves and large $V_{TH}$ modulation from −3.3 to +2 V when $V_{TG}$ is swept from +2 to −4 V (see Supplementary Figure S8a). Despite the large $V_{TH}$ modulation, the $I_{ON}/I_{OFF}$ remains nearly insensitive to changes in $V_{TG}$ (Figure S8b). To compare the gate tuning efficiency between BG and TG, the TG transfer characteristics at various $V_{BG}$ values were also extracted and are shown in Figure 2c. These transfer curves show a more sub-linear behavior, as compared with those of BG mode. In other words, the ON-state $I_D$ under TG mode is easy to saturate, and $V_{TH}$ shifts from −3.3 to +1.3 V when $V_{BG}$ drops from +2 to −4 V (see Supplementary Figure S9a). This modulation of $V_{TH}$ for the TG is smaller than what was observed in the transfer curves for the BG. Furthermore, the $I_{ON}/I_{OFF}$ values for the TG show a strong dependence on $V_{BG}$ (Figure S9b). In order to further compare the difference between BG and TG mode, the field-effect mobility $\mu_{FE}$ and SS values are extracted from Figure 2b-c, and shown in Figure 2d. $\mu_{FE}$ is calculated by the formula of $\mu_{FE} = \frac{dI_D}{dV_G} \times \frac{L}{WC_{ox}V_D}$,[31] where $L$ and $W$ are the channel length and width, respectively, $C_{ox}$ is the capacitance of the gate oxide. It is noteworthy that $\mu_{FE}$ extracted from the BG mode is nearly independent of $V_{TG}$ (about 18 cm$^2$/V·s); nevertheless, the $\mu_{FE}$ extracted from the TG mode increases from 1 to 25 cm$^2$/V·s as $V_{BG}$ increases from -4 to 1 V, mainly because $V_{BG}$ can also modulate $R_c$.[32-33]

The DG structure also influences the value of *SS*. We then perform further analysis by focusing on the contact regions, as shown in Figure 2e-f. For the case of BG operation, when $V_{TG}$<0, the top layer of MoS$_2$ is depleted and the electron injection path can be either from point 'A' to 'B' or from 'C' to 'B', then into the MoS$_2$ channel (Figure 2e left). This also results in an extra 'plateau' shown in the upper-right of Figure 2e, and the corresponding equivalent circuit diagram is shown in the lower-right of Figure 2e. For the case of TG operation, when $V_{BG}$<0, most electrons are injected from point 'A' to the MoS$_2$ channel to form an optimized path with lowest resistance (Figure 2f left), similar to that of a conventional MOSFET. The corresponding energy band diagram and equivalent circuit diagram of TG mode are shown in the right of Figure 2f. In this case, the top contacted source/drain electrodes fix the Fermi level of the MoS$_2$ and screen the electrical field from the TG,[34] so *SS* only depends on the TG control to the channel MoS$_2$, leading to a smaller *SS* compared with that of BG mode. It is indeed confirmed in Figure 2d, where the *SS* extracted from the TG mode (*SS*~100 mV/dec) is smaller than that of the BG mode (*SS*~130 mV/dec), when the control $V_G$ ($V_{BG}$ in the case of TG mode, and vice versa) is negatively applied (from -4 to -2 V). However, when the control $V_G$ is large enough to turn on the opposite surface of the channel, it provides an extra path for the current and thus *SS* starts to increase. In Figure 2d, it is also noteworthy that the degradation of *SS* for TG mode begins when $V_{BG}$ ~ -2.0 V, earlier than that of the BG mode ($V_{TG}$ ~ -1.0 V), which is mainly due to the $V_{BG}$ dependent $R_c$, as discussed above. Therefore, we can conclude that the BG mode provides a larger working current which is suitable for high-speed application; while the TG mode provides a steep switching for low power electronics.

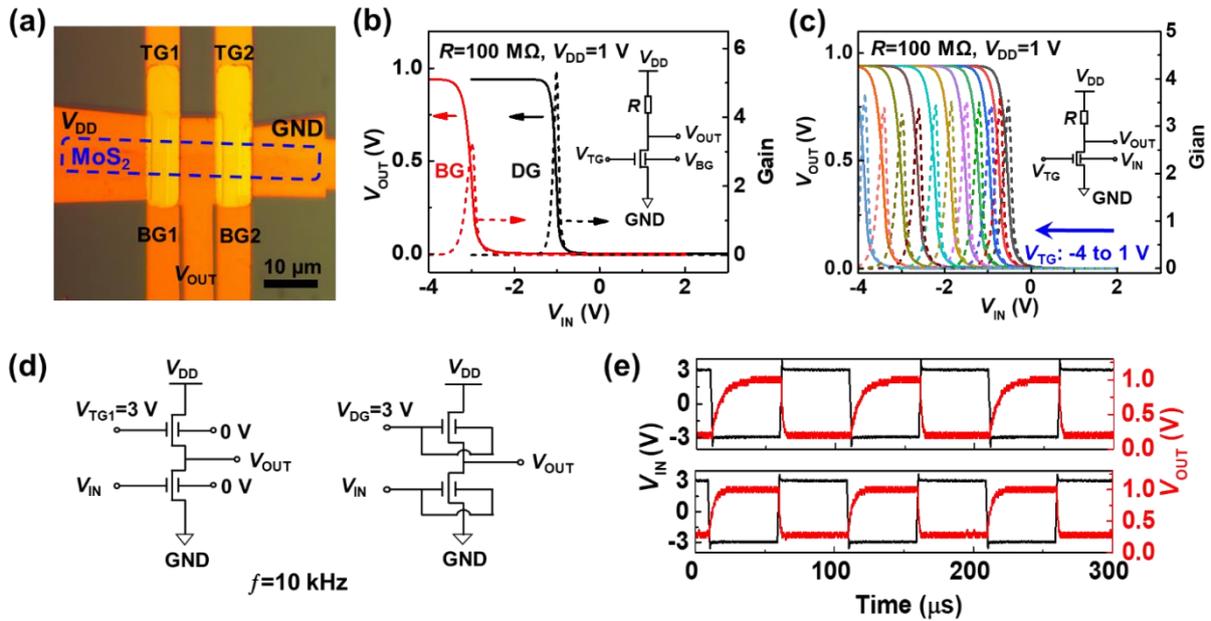

**Figure 3.** (a) An optical microscopic image of an as-fabricated integrated inverter circuit based on two monolayer MoS$_2$ DG-FETs connected in series. Scale bar is 10 μm. (b) Voltage transfer characteristics $V_{OUT}$-$V_{IN}$ (solid line) for the DG MoS$_2$ inverter in the BG and DG modes. The corresponding dependence of the inverter gain on $V_{IN}$ is also shown using short dashed lines. (c) Voltage transfer (solid line) and gain (short dashed lined) characteristics of the DG MoS$_2$ inverter modulated by varying $V_{TG}$. (d) The schematic electrical connection of the logic inverter in the SG (left) and DG modes (right). (e) Comparison of pulse response in the SG (upper panel) and DG (lower panel) modes. The black and red lines show the input and output signals, respectively.

Based on these advantages of the DG architecture for use with MoS$_2$, we proceed to pursue its practical application in logic circuits by fabricating a logic inverter. One MoS$_2$ DG-FET was used as a switch element, as shown in Figure 3a. To illustrate the $V_{OUT}$-$V_{IN}$ voltage transfer characteristics (VTC) in the BG and DG modes, a 100 MΩ resistor load was used for a simple comparison, as shown in Fig. 3b. When the logic inverter operates in the BG mode, i.e., $V_{BG}$ acts as the input voltage and $V_{TG}$ = 0 V, the switching threshold voltage is −3 V and the inverter gain (reciprocal of d$V_{OUT}$/d$V_{IN}$) is approximately 3. Meanwhile, the switching threshold voltage in the DG mode increases up to −1 V and the gain nearly doubles (~5.3), which indicates that better performance can be realized in the DG mode. We also explore the inverter VTC in the DG mode by tuning the driving voltage $V_{DD}$ from 0.5 to 3 V (see Supplementary Figure S10). Larger gain (~10) is obtained when $V_{DD}$ increases to 3 V. The $V_{TG}$-dependent output characteristics are also shown in Figure 3c. The VTC transition voltage $V_T$ of the inverter can be tuned from −0.5 to −3.8 V with stable gain (~ 3.5) as $V_{TG}$ increases from −4 to +1 V, indicating that $V_T$ can be controlled separately using one gate in the DG MoS$_2$ inverter, providing more flexibility in applications requiring low power consumption.

We also tested a depletion-load inverter based on two MoS$_2$ DG-FETs connected in series, as shown in Figure 3d. The response from the inverter circuit was measured with $V_{DD}$=1 V while driven with a 10 kHz square wave input signal in SG and DG modes, as shown in Figure 3e. Although the average propagation delay time ($\tau_P$) in the DG mode slightly exceeds 1.5 μs, which is rather limited due to parasitic capacitance from the D/S region, it is still much greater than the $\tau_P$ value for the SG inverter (7.5 μs), confirming that the DG structure is better suited for realizing high-speed applications. For DRAM applications, the OFF state and leakage currents affect the memory retention time which is an important factor for low-power operation. Our MoS$_2$ DG-FET can be further used in memory devices due to its excellent electrostatic control of the channel current. For a preliminary demonstration, a 1T1C DRAM unit cell was fabricated on a sapphire substrate, as shown in Figure 4a. In this circuit, a metal-insulator-metal (MIM) capacitor with a total capacitance of 102 pF was formed and integrated with a 3-nm-thick MoS$_2$ DG-FET (see Supplementary Figure S11 for more details). The response from the 1T1C cell was measured using an AC measurement technique (see Methods and Supplementary Figure S12) similar to that used in conventional DRAM arrays.[35-36] We then focus on the retention time, which determines the refresh rate and DRAM power consumption. Figure 4c and f show schematic circuits for the TG and DG modes, respectively. The corresponding pulse sequence was measured with an oscilloscope, and the pulse sequence is shown in Figure 4d-e and Figure 4g-h, respectively. During the write operation, one electrode was connected to the MoS$_2$ DG-FET and acted as the bit line (BL) while $V_{BL}$ switched from 0 to 1 V. This action simultaneously triggers the word line (WL) voltage $V_{WL}$, which is applied to either the TG/BG or DG electrode. The applied $V_{WL}$ bias switches from −3 to +3 V, where −3 V is the bias in the retention (hold) mode for the memory cell. During the write operation, the MoS$_2$ DG-FET is in the ON-state, thus the capacitor is charged by $V_{BL}$. The write rate is faster under the DG mode due to better channel control. To read the data, the voltage drop on the storage-node capacitor is compared with one-half of $V_{BL}$ (0.5 V in this case), and the amount of stored charge is determined by the sign and value of the current through the capacitor. While the capacitor discharges, the read operation results in a negative current spike, i.e., the current changes direction. This is observed in both Figure 4d and g when a hold time $t_{HOLD}$ = 10 ms is applied. However, when $t_{HOLD}$ ~ 700 ms, the sign of the current remains the same in the TG mode (Figure 4e). This indicates that slow charge leakage from the storage node primarily occurs because $I_{ON}/I_{OFF}$ for the

MoS$_2$-FET in the TG mode is limited, thus the voltage across the capacitor drops below 0.5 V applied to the bit line. In the DG mode, thanks to the superior electrostatic control over the channel current, the subthreshold leakage is much smaller than that in the TG mode, leading to a longer retention time (Figure 4h).

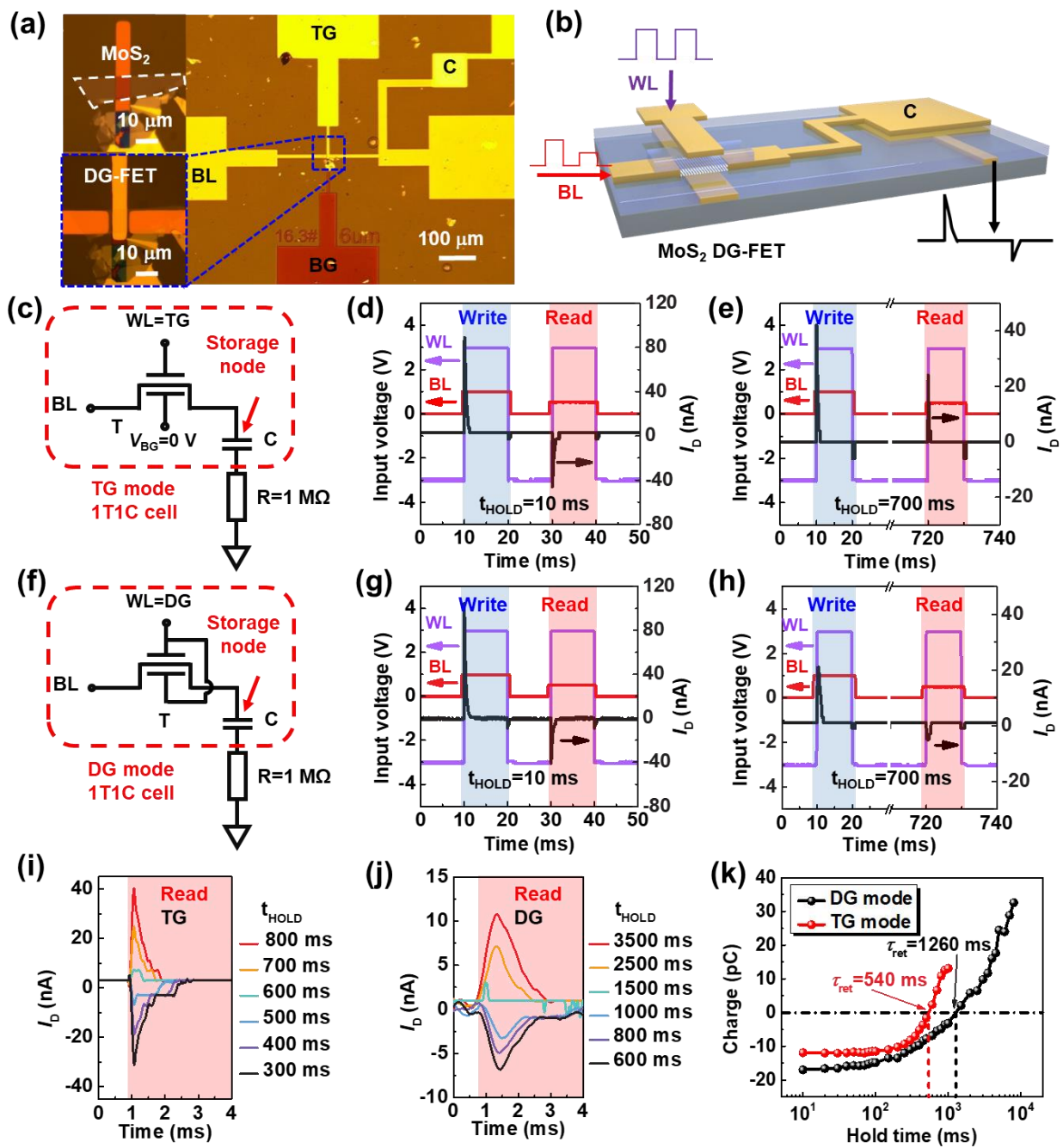

**Figure 4.** A 1T1C DRAM unit cell based on one MoS$_2$ DG-FET and one parallel-plate capacitor. (a) Optical micrograph of a 1T1C cell consisting of a ~3 nm MoS$_2$ DG-FET and a MIM capacitor on a sapphire substrate. The upper left image shows the fabrication step after a MoS$_2$ flake is transferred to the buried BG. (b) Schematic illustration of the operation of a DRAM cell with input and output signals. (c) Circuit of the DRAM cell in the TG mode. The internal resistance of the oscilloscope R is not part of the memory cell and is used to read the current flow during reading and writing operations. (d) DRAM operation with 10 ms hold time. The red and purple waves are applied to the bit and word lines, respectively, and the output current (extracted from the voltage across R) is shown in the black curve. (e) The same measurements as those in (d), but with 700 ms hold time. After such delay, the readout current has the same sign as that of the write pulse, indicating that the stored charge has been lost. (f) DRAM cell in the DG mode. (g-h) Corresponding memory operation with 10 ms and 700 ms hold times, respectively. (i) Readout current in the TG mode with hold time varying 300 to 800 ms in 100 ms increments. (j) Readout current in the DG mode with hold time varying from 600 to 3500 ms. (k) Calculated retained charge as a function of hold time in the TG (red) and DG modes (black) in our DRAM cell.

To further characterize the retention time $\tau_{ret}$, automated measurements were gathered, where write and read pulses were applied in succession while the hold time (i.e., the interval between write and read pulses) was varied. Figure 4i shows the current evolution measured during a read operation in the TG mode, clearly illustrating that the current changes direction when $\tau_{ret}$ ranges from 500 to 600 ms. However, according to Figure 4j, $\tau_{ret}$ for the DG mode lasts more than 1000 ms. The remaining amount of charge $Q_{READ}$ at the start of a read operation is calculated by integrating the current from the beginning to the end of a read cycle over time, and the retention time $\tau_{ret}$ is defined as the interval time over which $Q_{READ} = 0$. Figure 4k shows $Q_{READ}$ as a function of $t_{HOLD}$ in the TG and DG modes. One can clearly see that the TG mode provides a larger $Q_{READ}$ value for a given $t_{HOLD}$ value. $\tau_{ret} \approx 540$ ms in the TG mode, which is more than two times of the reference ($\tau_{ret} = 251$ ms),[37] while $\tau_{ret} \approx 1260$ ms in the DG mode. This $\tau_{ret}$ in the DG mode is approximately 20 times longer than the value in a silicon-based DRAM cell.[38] These results suggest that a DG MoS$_2$-FET has a potential application for memory devices.

These samples based on exfoliated MoS$_2$ sheets exhibit excellent device performance. However,

devices based on wafer-scale homogeneous films are highly desired for use in practical applications. Therefore, a batch-fabricated isolated 81 MoS$_2$ DG-FETs with $L$=10 μm and $W$=40 μm were also obtained based on CVD-grown MoS$_2$ film to demonstrate large-scale integration of our DG architecture for 2D-TMDs, as shown in Figure 5a. The wafer-scale vacuum stacking technique was used to form a desired number of ML-MoS$_2$ layers, which were transferred to the target substrate.[39-40] The synthesis and fabrication details are illustrated in Supplementary Figure S13-S15. Figure 5b shows measured BG, TG, and DG mode transfer characteristics based on the wafer-scale monolayer and transferred bilayer MoS$_2$ films (output curves see Figure S16). The cumulative transfer curves for 8 monolayer and 7 bilayer MoS$_2$ DG-FETs are displayed in Figure S17. The device uniformity is unsatisfactory, which is probably due to the inhomogeneity of the MoS$_2$ film and the contamination introduced by the film transfer process. In order to compare the performance of monolayer and bilayer MoS$_2$ DG-FETs, we present statistical comparison of $I_{ON}$, $I_{ON}/I_{OFF}$, $SS$ and $\mu_{FE}$ of these devices, as shown in Figure 5c and d. Similar to the results from the exfoliated samples, smaller $SS$ and larger $I_{ON}$ and $I_{ON}/I_{OFF}$ were achieved under the DG mode than those under the SG mode. In addition, average $I_{ON}$ of bilayer MoS$_2$ DG-FETs in DG mode is about 9 times higher than that of monolayer, and $I_{ON}/I_{OFF}$ of bilayer MoS$_2$ DG-FETs in DG mode is 2.2×10$^6$, even larger than that of monolayer (1.1×10$^6$). Both bilayer and monolayer devices show similar $SS$ (446~475 mV/dec), but the $\mu_{FE}$ of bilayer devices under DG mode is much larger than that of monolayer. More importantly, current measurements from wafer-scale bilayer MoS$_2$ films are already competitive to those from the exfoliated samples, while there is still significant room for improving the quality of CVD MoS$_2$, which shows great potential of our DG architecture to be used in practical device applications of 2D materials.

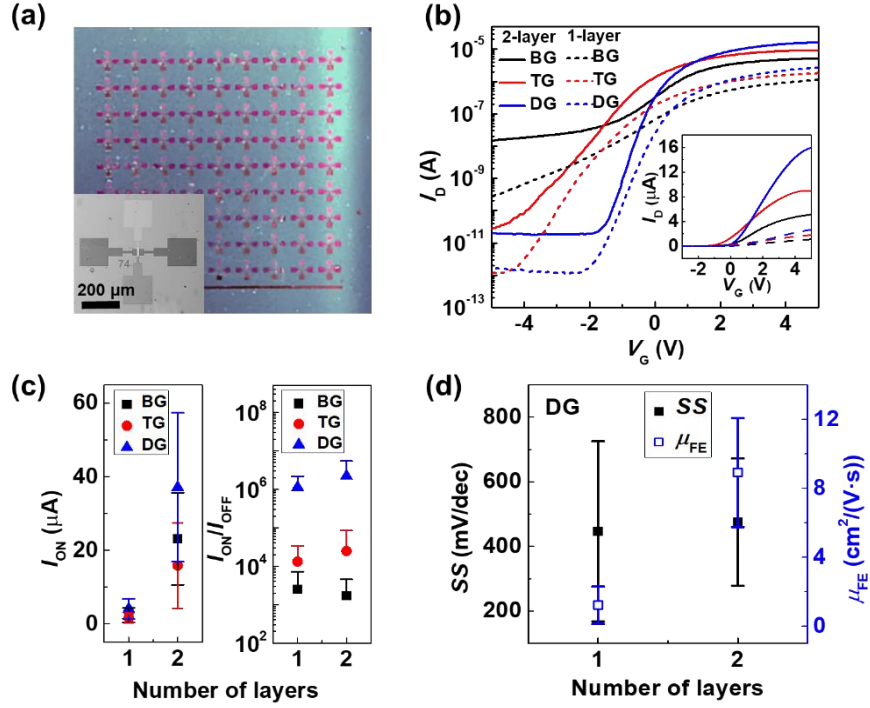

**Figure 5.** (a) Photograph of a die with isolated 81 DG-FETs based on a CVD grown $MoS_2$ bilayer film. The inset shows a magnified optical microscope image of a single $MoS_2$ DG-FET. (b) Room temperature transfer characteristics as functions of $V_{BG}$, $V_{TG}$, and $V_{DG}$ with $V_D = 0.1$ V for a typical monolayer and bilayer $MoS_2$ DG-FETs shown in (a). (c) Comparison of $I_{ON}$ (left), $I_{ON}/I_{OFF}$ (right) for 7~8 DG-FETs based on monolayer and bilayer $MoS_2$. (d) $SS$ and $\mu_{FE}$ of the monolayer and bilayer $MoS_2$ DG-FETs operating in DG mode.

## 3. CONCLUSIONS

We extensively investigated a symmetric DG architecture for $MoS_2$ FETs. Compared with the SG $MoS_2$ FET, the primary advantages of the DG structure can be summarized as follows. First, such a structure provides independent control of $V_{TH}$ by separately tuning BG and TG. Second, a large output current was observed for $MoS_2$ films with thickness ranging from 1 to 5 nm, and $SS$ is approximately <100 mV/dec with a large $I_{ON}/I_{OFF}$ ratio of >$10^7$; these values are competitive to the best SOI devices. Although such symmetric DG structure leads to the increment of complexity of the device integration and integrated circuit, it illustrates a new paradigm for 2D-TMDs based devices as it takes advantage of their ultrathin nature and provides an extra gate for flexible control of device performance and circuit operation. Finally to demonstrate the low power device application, a logic inverter with tunable gain and $V_T$, and a 1T1C DRAM cell with ultra-long retention time were fabricated based on the MoS$_2$ DG-FETs. Such device architecture can also be extended to other 2D materials.

## 4. EXPERIMENTAL SECTION

**Fabrication of the MoS$_2$ DG-FET.** The device fabrication begins by patterning a local bottom gate by photolithography on a heavily p-type doped Si substrate with 300 nm SiO$_2$ or sapphire substrate. Then CF$_4$+Ar plasma etching is used to obtain a 30 nm deep trench. 5 nm/25 nm of Ti/Au are then deposited by e-beam evaporation as a bottom gate. Subsequently, the high-k dielectric HfO$_2$ on the bottom gate is deposited by atomic layer deposition (ALD, MNT-100-4). During the ALD process, tetrakis (ethyl-methylamido) hafnium reacts at 180 °C with H$_2$O to form 15 nm HfO$_2$. MoS$_2$ flakes were first mechanically exfoliated from a 300 nm SiO$_2$/Si substrate by the classical scotch-tape technique[41] and then dry transferred to the as-fabricated bottom gate. 5 nm Ti/30 nm Au contacts are then defined on the MoS$_2$ sheets via electron beam lithography, evaporation and lift off. After lift-off in acetone, the device was annealed at 200 °C in a high vacuum furnace for 2 h to remove resist residue and to decrease contact resistance. And 2 nm Y layer was deposited by e-beam evaporation and then oxidized in oxygen furnace as a seed layer for subsequent ALD deposition of 15 nm HfO$_2$ and 5 nm Ti/40 nm Au of gate electrode onto the channel to form the top gate.

**Characterization and Electrical Measurements.** The thickness of MoS$_2$ is measured using an AFM (Dimension Edge, Bruker, USA) and Raman spectra (inVia Ranman Microscope, Renishaw, UK). All of the device, inverter and memory cell characterization are performed under air at room temperature. Capacitance-voltage characteristics are measured with a KEYSIGHT E4990A Impedance Analyzer.

Electrical characterizations are carried out with current-voltage measurements (Agilent, Semiconductor Characterization System B1500a).

**Fabrication and Characterization of the DRAM Cell.** In order to avoid some unexpected coupling capacitance, there is a non-overlapped region between the gate and the channel region. The *L* and *W* of this MoS$_2$ DG-FET are 8 μm and 11 μm, respectively. The capacitor has an area of 100×100 μm$^2$, the HfO$_2$ thickness is 15 nm and the dielectric constant is 17 for HfO$_2$, thus the total capacitance is 102 pF. A multichannel pulse generator is used for data writing and reading. The voltage pulses on BL and WL are synchronized so that $V_{BL}$ is raised 500 μs earlier, and decays 500 μs later than the $V_{WL}$. The duration of the WL pulse is long enough to ensure that the capacitor is fully charged. For the pulsed AC measurements, a KEYSIGHT 33600A series waveform generator is utilized. Both the input and output waveform data are monitored with a RIGOL DS1054Z digital oscilloscope. All the testing channels and the circuit shared a common ground terminal.

## ASSOCIATED CONTENT

**Supporting Information**

Experimental methods and supporting characterization data.

## AUTHOR INFORMATION

**Corresponding Author**

*E-mail: baowz@fudan.edu.cn (W.B.).

*E-mail: ychai@polyu.edu.hk (Y.C.).

*E-mail: pengzhou@fudan.edu.cn (P.Z.).

**Author Contributions**

Y.X., Y.C. and W.B. conceived the experiments, F.L., Z.G. and Y.W. designed and conducted the experiments. S.Z., Y.S. and H.T. analyzed the results, Z.X., P.Z. and D.W.Z provided the experiment conditions, A.R., J.W., M.S.F. and X.J. helped perform the analysis with constructive discussions. The manuscript was written through contributions of all authors. All authors have given approval to the final version of the manuscript. ‡These authors contributed equally.

**Notes**

The authors declare no competing financial interest.

ACKNOWLEDGMENTS

This work was supported by the National Key Research and Development Program of China (2016YFA0203900, 2018YFA0306101) and Natural Science Foundation of China (Grant No: 61874154). M. Fuhrer acknowledges support from the Australian Research Council under grant DP150103837.

**Figure Captions**

**Figure 1.** (a) Three-dimensional schematic of a MoS$_2$ DG-FET. (b) Optical image of an as-fabricated monolayer MoS$_2$ DG-FET. The scale bar is 10 μm. (c-d) Capacitance-voltage curves (below) and measurement connection (above) in the TG- and BG modes. (e) $I_D$-$V_D$ output curves with forward and backward sweep for various applied $V_{BG}$ (left), $V_{TG}$ (middle), and $V_{DG}$ (right) values (all range from −3 to 4 V with 1 V increments). The thickness of MoS$_2$ is nearly 1 nm. (f) $I_D$-$V_G$ transfer curves by forward and backward sweeping $V_{BG}$, $V_{TG}$, and $V_{DG}$ with $V_D$=0.1 V. (g) *SS* extracted from the BG (black hollow squares), TG (red hollow circles), and DG (blue solid circles) transfer curves for MoS$_2$ with different thicknesses. The dashed line is 60 mV/dec and the blue line is the fitting result. (h) Dependence of $I_{ON}/I_{OFF}$ in DG-FETs on the thickness of MoS$_2$.

**Figure 2.** Electrical characteristics of a monolayer MoS$_2$ DG-FET. (a) 2D mapping of $I_D$ as functions of $V_{TG}$ and $V_{BG}$ at a constant $V_D$ = 0.1 V. (b) BG Transfer characteristics with $V_{TG}$ ranging from −4 to 2 V in 1 V increments. (c) TG transfer characteristics with $V_{BG}$ ranging from −4 to 0 V in 0.5 V increments. (d) The extracted $\mu_{FE}$ (black symbols) and *SS* (red symbols) operated under the two modes. Square symbols are extracted from curves in (b), corresponding to the BG mode under various $V_{TG}$, and circular symbols are extracted from curves in (c), corresponding to the TG mode under various $V_{BG}$. (e) A cross-sectional view of the contact region when operated under BG mode with $V_{TG}$<0 V (left), and the corresponding energy band diagrams for BG modulation (upper right) and equivalent circuit diagram (lower right). (f) A cross-sectional view of the contact region when operated under TG mode with $V_{BG}$<0 V (left), and the corresponding energy band diagrams for TG modulation (upper right) and equivalent circuit diagram (lower right). The solid and dashed lines in the band diagrams correspond to the negative and positive gate voltages, respectively.

**Figure 3.** (a) An optical microscopic image of an as-fabricated integrated inverter circuit based on two monolayer MoS$_2$ DG-FETs connected in series. Scale bar is 10 μm. (b) Voltage transfer characteristics $V_{OUT}$-$V_{IN}$ (solid line) for the DG MoS$_2$ inverter in the BG and DG modes. The corresponding dependence of the inverter gain on $V_{IN}$ is also shown using short dashed lines. (c) Voltage transfer (solid line) and gain (short dashed lined) characteristics of the DG MoS$_2$ inverter modulated by varying $V_{TG}$. (d) The schematic electrical connection of the logic inverter in the SG (left) and DG modes (right). (e) Comparison of pulse response in the SG (upper panel) and DG (lower panel) modes. The black and red lines show the input and output signals, respectively.

**Figure 4.** A 1T1C DRAM unit cell based on one MoS$_2$ DG-FET and one parallel-plate capacitor. (a) Optical micrograph of a 1T1C cell consisting of a ~3 nm MoS$_2$ DG-FET and a MIM capacitor on a sapphire substrate. The upper left image shows the fabrication step after a MoS$_2$ flake is transferred to the buried BG. (b) Schematic illustration of the operation of a DRAM cell with input and output signals. (c) Circuit of the DRAM cell in the TG mode. The internal resistance of the oscilloscope R is not part of the memory cell and is used to read the current flow during reading and writing operations. (d) DRAM operation with 10 ms hold time. The red and purple waves are applied to the bit and word lines, respectively, and the output current (extracted from the voltage across R) is shown in the black curve. (e) The same measurements as those in (d), but with 700 ms hold time. After such delay, the readout current has the same sign as that of the write pulse, indicating that the stored charge has been lost. (f) DRAM cell in the DG mode. (g-h) Corresponding memory operation with 10 ms and 700 ms hold times, respectively. (i) Readout current in the TG mode with hold time varying 300 to 800 ms in 100 ms increments. (j) Readout current in the DG mode with hold time varying from 600 to 3500 ms. (k) Calculated retained charge as a function of hold time in the TG (red) and DG modes (black) in our DRAM cell.

**Figure 5.** (a) Photograph of a die with isolated 81 DG-FETs based on a CVD grown MoS$_2$ bilayer film. The inset shows a magnified optical microscope image of a single MoS$_2$ DG-FET. (b) Room temperature transfer characteristics as functions of $V_{BG}$, $V_{TG}$, and $V_{DG}$ with $V_D$ = 0.1 V for a typical monolayer and bilayer MoS$_2$ DG-FETs shown in (a). (c) Comparison of $I_{ON}$ (left), $I_{ON}/I_{OFF}$ (right) for 7~8 DG-FETs based on monolayer and bilayer MoS$_2$. (d) $SS$ and $\mu_{FE}$ of the monolayer and bilayer MoS$_2$ DG-FETs operating in DG mode.

**TOC**

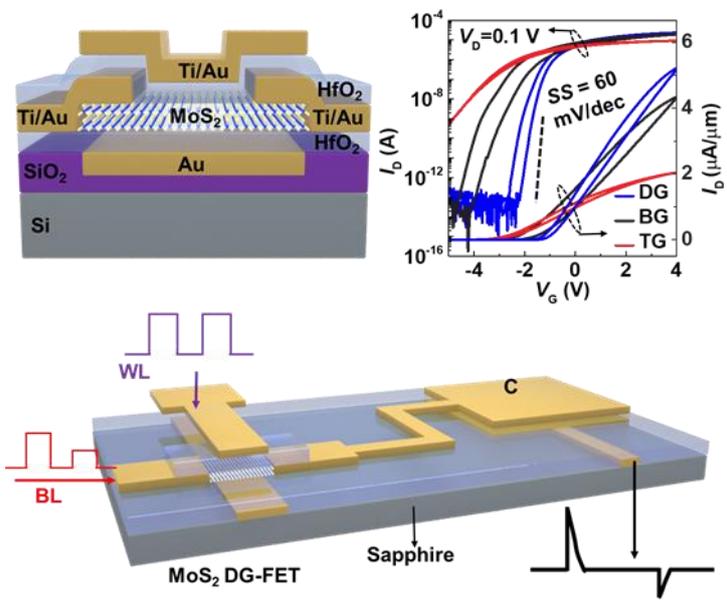